%Paper: astro-ph/9408063
%From: "Home Page URL: http://enemy.gsfc.nasa.gov/htmltest/rjn.html"
%%<NEMIROFF@grovx1.gsfc.nasa.gov>
%Date: Wed, 17 Aug 1994 17:46:53 -0400 (EDT)

% This is plain TEX, version 3.0, without macros.
% Postscript text, tables and figures available through RJN's URL:
% http://enemy.gsfc.nasa.gov/htmltest/rjn.html

\magnification=1200
\baselineskip=14pt

\def\cl{\centerline}

\ \

\bigskip\bigskip
\bigskip\bigskip
\bigskip\bigskip

\cl{\bf DURATION DISTRIBUTIONS OF BRIGHT AND DIM }
\cl{\bf BATSE GAMMA-RAY BURSTS }

\bigskip\bigskip
\bigskip\bigskip

\cl{ J.P. Norris$^{(1)}$, J.T. Bonnell$^{(1),(2)}$, R.J. Nemiroff$^{(1),(3)}$,
}
\cl{ J. D. Scargle$^{(4)}$, C. Kouveliotou$^{(2),(5)}$, }
\cl{ W.S. Paciesas$^{(5),(6)}$, C. A. Meegan$^{(5)}$, G.J. Fishman$^{(5)}$ }

\bigskip\bigskip
\bigskip\bigskip

\cl{\it $^{(1)}$ NASA Goddard Space Flight Center, Greenbelt, MD 20771 }
\cl{\it $^{(2)}$ Universities Space Research Association }
\cl{\it $^{(3)}$ George Mason University, CSI Institute, Fairfax, VA 22030 }
\cl{\it $^{(4)}$ NASA Ames Research Center, Moffet Field, CA 94035 }
\cl{\it $^{(5)}$ NASA Marshall Space Flight Center, Huntsville, AL 35812 }
\cl{\it $^{(6)}$ University of Alabama, Huntsville, AL 35899 }

\bigskip\bigskip
\bigskip\bigskip
\bigskip\bigskip
\bigskip\bigskip
\bigskip\bigskip
\bigskip\bigskip

\cl{ In press: The Astrophysical Journal (Letters) }
\cl{ Submitted to The Astrophysical Journal on June 13, 1994 }
\cl{ Revised:  July 29, 1994 }

\vfill\eject
\baselineskip=18pt

\ \
\bigskip\bigskip

\cl{\bf ABSTRACT }

\bigskip

We have measured the T90 and T50 durations of bright and dim gamma-ray
bursts (GRBs) detected by the Compton  Gamma Ray Observatory's Burst and
Transient Source Experiment (BATSE).  The T90 (T50) duration is defined as
the interval over which 5\% (25\%) to 95\% (75\%) of the burst counts
accumulate.  Out of 775 bursts observed by BATSE, 159 bursts were analyzed;
bursts with durations shorter than 1.5 s were excluded.  A
Kolmogorov-Smirnov test yields a probability of 6 x $10^{-5}$ that the T50
durations of the dim and bright samples are drawn from the same parent
population.  We find that the centroid and extent of the duration
distribution for the dim sample are scaled by approximately a factor of two
relative to those of the bright sample.  The measured time dilation factor
is not sensitive to choice of energy band.  These results are
quantitatively consistent with previous tests for time dilation in a
smaller sample of BATSE bursts. The dimmer bursts, if cosmological, would
lie at redshifts of order two.

\bigskip

\noindent {\it Subject headings:} cosmology:  theory - gamma rays:  bursts

\vfill\eject

\cl{\bf 1. INTRODUCTION }

\bigskip

Over one thousand gamma-ray bursts (GRBs) have been detected by the Burst
and Transient Source Experiment (BATSE) on the Compton Observatory.
Analysis of refined positions for three-fourths of the bursts still
indicates no significant deviation from isotropy, whereas their
number-intensity relation implies an effective limit to the depth of the
spatial distribution (Meegan et al. 1994).  Norris et al. (1994a) and Davis
et al. (1994) performed several tests, and found a time dilation factor of
approximately two between bright and dim bursts.  Using the same brightness
groupings (see Table 1), Nemiroff et al. (1994) report a significant
difference in spectral hardness ratios as a function of time, possibly
consistent with cosmological redshifts.  These findings, consistent with
predictions by Paczynski (1992) and Piran (1992), lend increasing support
to the cosmological hypothesis, especially in light of interpretations of
the number- intensity relation which are consistent with redshifts of order
1 - 2 for the dimmer bursts (Mao \& Paczynski 1992; Wickramasinghe et al.
1993; Fenimore et al. 1993).  (Also, in the general context of cosmological
time dilation, see Noerdlinger [1966] and McCrea [1972].)  However,
analysis of a different sample of BATSE bursts (Mitrofanov et al. 1994),
more uniformly spread in intensity and not representing the dim and bright
samples of Norris et al. (1994a) did not find a significant time dilation
effect.

In this paper we adopt an objective definition of the "duration" of a
burst, and then study the dependence of this quantity on brightness, using
the same peak intensity groupings as in Norris et al. (1994a).  We
introduce modifications to the procedure used by Kouveliotou et al. (1993)
to obtain T90 and T50 duration estimates.  The modifications realize an
automated procedure for estimating burst durations, while using a noise
reduction technique to reduce variances which arise from finite counting
statistics and heterogeneous burst profiles.  Brightness selection effects
are avoided by rescaling all bursts to fiducial levels of peak intensity
and noise bias; the analysis is also performed without such equalization of
the signal-to-noise ($S/N$) level.  We find that the centroid and extent of
the duration distribution for the dim sample are scaled by approximately a
factor of two relative to those of the bright sample, in good agreement
with the time dilation factor found from multi-resolution tests (Norris et
al. 1994a), and consistent with cosmic time dilation.

\bigskip

\cl{\bf 2. ESTIMATION OF DURATIONS }

We analyzed burst samples in three peak intensity ranges (described in
Table 1), using data from BATSE's Large Area Detectors (LADs).  The burst
duration distribution is bimodal, the modes of the sample in the BATSE
first catalog being ~ 330 ms and ~ 25 s, as shown by Kouveliotou et al.
(1993).  To avoid any possible trigger selection bias, we elected to study
only the long mode of the duration distribution and therefore excluded
bursts shorter than $\sim$ 1.5 s. Thus, of a total sample of 775 bursts
detected by BATSE, 159 were analyzed - those bursts longer than 1.5 s and
within the intensity ranges in Table 1.  Following Kouveliotou et al., we
utilized count rates summed over the four LAD channels ($>$ 25 keV); we
also performed the analysis for channels 2 - 3 (58 - 320 keV) and for
channel 3 (115 - 320 keV).  Peak intensities were determined on a 256 ms
timescale, following a procedure similar to that described in Norris et al.
(1994a), with one salient difference that allowed examination of time
intervals of at least 340 s:  we utilized 100 s of pretrigger data with
1.024 s resolution in addition to the high resolution (64 ms) data
contained in the burst record.  Each of the 1.024 s pretrigger samples was
divided into sixteen 64 ms samples, with the counts randomly distributed in
time at the finer resolution (thus preserving Poisson statistics); the
pretrigger and burst record data were then concatenated in correct temporal
registration. The concatenated time profiles were binned to 512 ms for
further analysis.

Visual examination of the time profiles in the three lowest LAD energy
channels (25 - 50 keV, 50 - 100 keV, 100 - 300 keV) yielded coarse
estimates of burst durations.  Pulse structures are often more easily
discernible at higher energy, whereas burst envelopes persist longer at
lower energy.  For bursts with apparent total durations shorter than 131 s
(2048 samples times 64 ms; $2n$ points required for analysis procedure),
backgrounds were fitted using a quadratic form.  Adequate background fits
necessitated fourth or fifth order polynomials for twelve longer bursts
(4096 samples used).  In order to eliminate brightness selection effects on
the duration measures, we diminished the background-subtracted profiles to
a canonical peak intensity, added a canonical constant background, and
equalized the Poisson counting rate.  The canonical peak intensity chosen
(1400 counts s$^{-1}$ was that of the dimmest burst used in the study.
This equalization procedure is detailed in Norris et al. (1994a).  However,
because there is a tradeoff between addressing brightness selection effects
and preserving original signal-to-noise levels, the entire analysis was
also performed without equalization of the $S/N$ level.

Kouveliotou et al. (1993) measure the T90 duration, defined to be the
interval between times when 5\% and 95\% of the burst counts (above
background) are accumulated.  Similarly, T50 is the interval between 25\%
and 75\% of the cumulative counts.  We also generate these measures, but
using an automated procedure, once backgrounds are fitted and subtracted.
Two features of our approach combine to reduce the effects of statistical
fluctuations.  First, each profile is passed through high and low pass
filter networks (Donoho 1993; Donoho \& Johnstone 1994).  The high pass
filter network is a Haar wavelet transform in dyadic steps (Daubechies
1992), followed by application of a (possibly different) threshold at each
timescale represented in the wavelet transform, and then an inverse wavelet
transform (see Norris et al. 1994a).  Thresholding on each timescale for
the high pass filter results essentially in removal of wavelet coefficients
which represent less than one sigma differences from point to point.  The
low pass network is the sequence of compressed time profiles, also in
dyadic steps.  The low pass filter also removes wavelets of the same
amplitude range, but measured against the ambient background level rather
than against the ensemble of wavelet amplitudes.  The efficacy of such high
and low pass filters is that statistical fluctuations about background
level are largely deleted, while significant burst structure is preserved.
Using simulations, we have demonstrated that this multi-resolution approach
preserves temporal shape (on timescales $>$ 128 ms) to relatively high
precision.

Second, the same low pass filter procedure was used as a ``floating
trigger" (much as the BATSE experiment itself has triggers on three
timescales), to determine lower and upper limits on the $t_0$ (onset) and
$t_{100}$ (cessation) times of the burst, respectively.  These limits for
the $t_0$ and $t_{100}$ points were determined within the 131 s intervals
for bursts with quadratic background fits, and within the 262 s intervals
for the twelve longer bursts.  The initial $t_0$ and $t_{100}$ points were
assigned the times of the last and first bins, respectively, of the burst
profile. Successively for each timescale, the $t_0$ ($t_{100}$) point was
decreased (increased) to the time of the earliest (latest) fluctuation of
at least one sigma significance above background.  The threshold was
reduced by the factor $1/2$ for each dyadic compression of the time series,
thereby maintaining a constant threshold level across all timescales.
Thus, long structures of lower intensity tend to be included as part of the
burst by this multi-resolution technique while shorter structures of higher
(but insignificant) intensity are ignored.  The floating trigger was
operated on timescales from 512 ms to 32 s.  To compensate for
discretization, the interval thus found was lengthened at start and end by
16 s (half the longest timescale) to yield the nominal $t_0$ and $t_{100}$
points.  Each time profile was then integrated between these times to
compute the {$t_5$,$t_{95}$} and {$t_{25}$, $t_{75}$ points. Note that the
first step, thresholding of the wavelet coefficients, considerably reduces
the random walk between the nominal $t_0$ and $t_{100}$ points and the actual
burst onset and cessation times, respectively.

Two important consistency checks were then performed.  Figure 1 compares
the T90 and T50 duration determinations for the bright bursts with original
signal-to-noise levels and without application of the high and low pass
filter procedure (dotted line) to the determinations with noise biases
rendered equal to that of the dimmest burst analyzed and with the filter
applied (solid line). For the noisy profiles, durations were determined
from the median of eleven realizations of the $S/N$ equalization procedure
(in the large majority of cases, the median was representative of most of
the eleven measurements).  The small differences between the two T90
distributions (Figure 1a) are consistent with statistical fluctuations.
More to the point, the centroids and widths of the two are practically
identical.  As T50 measures the core of the burst, where the intensity
levels tend to be well above background, one expects this measure to be
less sensitive to the S/N equalization procedure.  Further, T50 should be
less affected by outlying structures.  Thus, regardless of the particular
measurement procedure employed, T50 estimates may be more reliable - in the
sense that brightness selection effects are less pronounced.  In fact, the
T50 distributions (Figure 1b) for the two approaches are in slightly better
agreement than the T90 distributions.  We also compared our T90's with
those published in the first BATSE catalog (Fishman et al. 1994).  For the
eighteen bright bursts which appear in the catalog, fourteen had comparable
determinations.  Durations for four bright bursts with outlying,
low-intensity structures of very low-significance were underestimated by
our technique.  We note that statistical fluctuations in intrinsically dim
bursts would often obliterate such features - thus the rationale for
performing the $S/N$ equalization procedure.  In fact, our determinations for
dim bursts are in closer agreement with those in the first BATSE catalog
than our determinations for bright bursts.

The estimated T90 and T50 duration distributions for the three intensity
samples are illustrated in Figures 2 and 3, with (Method 1) and without
(Method 2) application of the $S/N$ equalization procedure, respectively.
Below, results from Method 1 are listed with those from Method 2 following
in parentheses.  For both methods, there is a clear shift of the two dim
samples to longer timescales with respect to the bright sample.  To
estimate if the duration distributions could have been drawn from the same
parent distribution, we performed a two-distribution Kolmogorov- Smirnov
(KS) probability analysis (Press et al. 1992).  The T90 durations for the
bright group were multiplied by a time dilation factor. The artificially
dilated durations were then statistically compared with the those of the
dim samples combined.  This was done for candidate time dilation factors
ranging from 1.0 to 4.0.  The results are shown in Figures 4a \& 4b for
Methods 1 and 2, respectively.  The duration distribution for the brights
is most similar to that of the dims when a time dilation factor of 1.9
(2.1) is applied - and the KS statistic indicates that the goodness of fit
is acceptable for both methods.  A time dilation factor of unity would
yield a KS statistic lower than that observed only once every 1500 (1500)
trials.  The horizontal line depicts the level above which there is at
least a 31.6\% chance that the brights and dims were drawn from the same
parent population.  Method 2 produces distributions of more comparable
logarithmic width (discussed below) than does Method 1.  This is reflected
in the fact that the maximum KS statistic, corresponding to the best time
dilation factor, is nearer unity for Method 2.  Figures 4a \& 4b also show
the KS statistic for the T50 duration distributions.  The best fit time
dilation factors were found to be 2.2 (2.4).  For the T50 measure a time
dilation factor of unity would yield a lower KS statistic than that found
only once in 16,000 (26,000) trials.

To estimate centroids and widths, we fitted a Gaussian form to the T90 and
T50 distributions, binned logarithmically with 5 bins per decade (fits with
10 bins per decade yielded practically identical results).  The results are
listed in Table 1 for the three brightness samples and for the two dim
groups combined.  A Gaussian form affords an adequate, but not optimal,
estimate of the moments of the duration distributions.  Whereas the KS test
tends to emphasize the central region of a distribution, a Gaussian fit
takes into account shape contributions from the wings.  Again, a clear
shift of approximately a factor of two is manifest between the centroids of
the bright and combined dim sample distributions.  For the T90 measure the
ratio of the centroids is 2.1 (2.4); for T50, the ratio is 2.2 (2.3).  For
both measures, considering the two dim samples separately yields somewhat
discrepant results:  via Method 1, T90 and T50 centroids for the dimmest
group are lower (33.4 s and 11.8 s, respectively) than for the dim group
(43.6 s and 15.4 s, respectively). Some possible explanations for these
disparate values are discussed in the next section.

To search for any trend in time-dilation factor with energy, we also
performed the analysis for channels 2 - 3 (58 - 320 keV) and for channel 3
(115 - 320 keV).  Fewer counts are available as fewer channels are
included, but an improvement in signal-to-noise level is realized by
excluding channel 4, in which the count rate is relatively low (emission is
often indiscernible in channel 4 for dim bursts).  Centroid and logarithmic
width determinations were comparable in all three energy ranges.  Table 2
summarizes the ratios of centroids (time-dilation factors) - bright sample
compared with combined dim and dimmest samples - for the several
combinations of energy band, method, and duration definition.  There is no
apparent trend with increasing energy.  The errors in time-dilation factor,
of order 0.25 unit, are comparable to the variations between methods and
between T90 and T50 measures.

Inspection of Figure 2 and Table 1 reveals that for Method 1, the bright
burst distribution is significantly narrower than any measure of the width
of the distribution of dim bursts. Logarithmic widths (Gaussian standard
deviation in logarithm of time) for the latter groups range from 0.41
$\pm$0.03 (in log [seconds]) to 0.54 $\pm$ 0.04, whereas for the bright
sample the values are 0.34 (0.35) $\pm$ 0.03 for T90 (T50). For Method 2
the difference in logarithmic widths is still present, but less significant
(see Figure 3) except when comparing the T50 measures for the bright and
dimmest samples. Note that for pure time dilation, neglecting all other
effects, the distributions should have the same widths in logarithm of
time.  Assume for sake of argument that Method 1 gives better estimations:

It would appear that the sample cutoff of 1.5 s cannot wholly account for
the narrower bright distribution since for the T90 measure, no durations
are determined to be less than 5 s.  It is possible that the measured
disparity in distribution width arises from finite sample size. We find
from redistributing the durations of only five bright bursts in the two
highest bins (10 - 25 s; see Figure 2) into the two adjacent bins, that the
logarithmic width broadens to ~ 0.40.  Also, for the optimum time-dilation
factors of 1.9 and 2.2, the KS test for Method 1 (see Figure 4) yields
probabilities of 0.56 (T90) and 0.63 (T50), respectively, that dim bursts
and time-dilated bright bursts are drawn from the same parent population.
For Method 2, the corresponding maximum probabilities are probabilities of
0.97 (T90) and 0.67 (T50).  We conclude that, within measurement errors,
the logarithmic widths of the bright and dim distributions are consistent
with pure time dilation.  Note, however, that other physical processes
intrinsic to the burst source may give rise to significant effects which
modify the relative shapes of the duration distributions.  In the next
section we briefly explore this possibility.

\bigskip

\cl{\bf 3. DISCUSSION }

We have estimated centroids and logarithmic widths of duration
distributions of bright and dim BATSE gamma-ray bursts.  The determinations
are in good agreement with previous multi- resolution tests which, when
calibrated with simulations, yielded an average time-dilation of 2.25
between bright and dim groups (Norris et al. 1994a).  This is in good
agreement with the time- dilation factor found for the T90 and T50 duration
measures, whether one considers the method utilizing original
signal-to-noise levels or the method of equalizing noise biases to yield
more reliable results.  With current sample sizes, Kolmogorov-Smirnov tests
allow the rejection of the hypothesis that the samples are drawn from the
same parent population at a confidence level of 6 x $10^{-5}$.  Thus, the
appearance that previous experiments (Mazets 1981; Norris et al. 1984;
Klebesadel 1990; Hurley 1991) preferentially detected bursts shorter than
BATSE bursts (in the duration range above 1.5 s) could be explained by the
fact that those experiments had less sensitivity and therefore sampled the
bright end of the number-intensity relation.  We suggest that attempts be
made to apply rigorously equivalent methods for duration measurement and to
intercalibrate instruments, and thereby increase the number of bright
bursts which may be compared with BATSE-detected dim bursts.

Recall that for both the T90 and T50 measures the distribution centroids of
the dimmest group are lower than for the dim group.  We discuss several
possible explanations.  Two possible systematic trends that might have been
introduced by the analysis method - which we believe to have negligible
effect on the overall shapes of the distributions - are background
subtraction and truncation of the analyzed ``burst interval" at 131 s or 262
s.  For progressively dimmer events, onset and cessation of emission
becomes more difficult to apprehend.  Thus in some dim bursts, we may have
selected background regions where in fact some very low level emission
occurs. Also, one dim burst in our sample is known to have a T90 duration
of $\sim$ 430 s - the additional emission comes prior to the 100 s of
pretrigger data which we utilize.  There are probably a few additional dim
bursts in which we have missed such precursor emission, but this cannot
have affected the bulk of durations for the dimmest group.  A third
explanation of the disparity in centroids for the dim and dimmest groups
requires the assumption of cosmological origin:  the dimmer bursts are more
redshifted, and therefore narrower structure from higher energy is shifted
into the BATSE LAD energy range, effectively shortening each pulse
structure (but, see below). Evaluation of this possibility will require
detailed comparison of observations and models. Finally, the difference may
be purely attributable to sample variations.

If bursters are at cosmological distances, a duration measure such as T90
or T50 may well prove to be as good or better a time dilation indicator as
multi-resolution measures of burst structure:  duration estimation for
entire bursts, which average $\sim$ 10 pulses (Norris et al. 1994b), should
be less subject to energy-dependent pulse width narrowing which would arise
from spectral energy redshift.  Indeed, by performing our analysis for
different energy bands, we have shown that there is no apparent trend of
duration time-dilation factor with energy.  We note however that, by
itself, duration does not address the necessary condition that all
structures and intervals between structures be time dilated by the same
factor.  New tests which measure pulse widths and intervals between pulses
do corroborate previous multi-resolution tests for time dilation (Davis
1994).

Assuming the bursters are at cosmological distances and following the
treatment of Wickramasinghe et al. (1993), we determine that the median
peak intensity (39,000 counts s$^{-1}$) of the bright sample would fall at a
redshift, z, of $\sim$ 0.3.  As discussed in Norris et al. (1994a), ``K-
corrections" for energy-dependent pulse width and spectral redistribution
arising from redshift must be applied, and modeling of intrinsic luminosity
functions performed to arrive at quantitative comparisons with theory.  But
if we take the duration dilation factor found in this study to be
relatively free of the spectral complications that must be considered for
shorter structures, then from $(1 + z_{dim})/(1 + z_{bright}) \sim 2.25$,
we can estimate for our dim sample that $z_{dim} \sim  2$.  This is
considerably higher than the redshift ($\approx$ 0.8) found for dim BATSE
bursts by combining the PVO and BATSE number-intensity relations, assuming
no source evolution with cosmic time (Fenimore et al. 1993).  Emslie \&
Horack (1994) have shown that only narrow (redshift-independent) luminosity
functions can be accommodated by the BATSE number-intensity relation unless
a finite cosmological constant is invoked and that z $\sim$ 1.25 would be
indicated for the dimmest BATSE bursts.  However, the acceptable parameter
space for the cosmological hypothesis is much wider if bursters evolve with
cosmic time either in luminosity or rate of occurrence, or both.  For
example, if $z_{dim} \sim 2$, then pure rate evolution with a redshift
dependence of $(1+z)^{0.6}$ provides an acceptable fit to the BATSE
number-intensity relation (Horack, Emslie \& Hartmann 1994).

Recently, Brainerd (1994) and Yi \& Mao (1994) have discussed the
complicating special relativistic (SR) effect of beaming, which is
necessary in many cosmological (or even Galactic halo) scenarios to account
for production and/or survival of high-energy gamma-rays (Dingus et al.
1994).  Brainerd showed that a spatially limited distribution of sources in
Euclidean space can exhibit an inverse correlation of intensity with
duration, and that beaming is one mechanism which would operate in that
context to produce such an average correlation.  However, Wijers \&
Paczynski (1994) show generally that for local (noncosmological) source
populations with distance-independent luminosity functions, the duration
distributions for high and low intensity burst samples will have different
cutoffs at short durations, but both distributions will extend to the same
cutoff at long durations.  This is in contradistinction to our result, in
which the distributions appear to exhibit a relative shift.  Wijers \&
Paczynski, using burst durations from the first BATSE catalog (Fishman et
al. 1994), find a result similar to ours, supporting a shift in
distributions rather than just a difference in the short duration regime.
Again, a shift with distribution shape preserved in logarithm of duration
is the expectation for pure cosmological time dilation.  Since detailed
models of duration distributions resulting from assumption of various
luminosity functions and spatial distributions is yet to be presented, it
is judicious to reserve judgment on the meaning of our results.

An important study, assuming trigger selection effects can be correctly
understood, will be analysis of bright and dim bursts with durations $<$
1.5 s.  Short bursts share the same peak intensity range as long bursts
(Kouveliotou et al. 1993), and may in fact be the same population (Norris
et al. 1994b).  Since pulses in short bursts are narrower than those in
long bursts, and since cosmological time dilation requires all intervals to
be stretched equally, additional constraints on beaming and cosmological
hypotheses should come from considering both short and long bursts.

We express our thanks to David Donoho for helpful suggestions.  JDS was
partially supported by the NASA Astrophysics Data Program.  The Second
BATSE Burst Catalog, is available electronically from the Compton
Observatory Science Support Center, at Decnet node GROSSC, user name
GRONEWS.

\bigskip

\cl{\bf REFERENCES }
{
\parindent=0pt
\hangindent=20pt
\baselineskip=18pt
\parskip=4pt

\bigskip

Brainerd, J.J. 1994, ApJ, 428, L1

Daubechies, I.  1992, Ten Lectures on Wavelets (Philadelphia:  Capital City
Press)

Davis, S.P.  1994, Ph.D. thesis

\hangindent=20pt
Davis, S.P., Norris, J.P., Kouveliotou, C., Fishman, G.J., Meegan, C.A., \&
Paciesas, W.S. 1994, in Gamma-Ray Bursts - Huntsville, AL 1993, AIP 307, in
press

\hangindent=20pt
Dingus, B., et al.  1994, in Gamma-Ray Bursts - Huntsville, AL 1993, AIP
307, in press

\hangindent=20pt
Donoho, D.L.  1993, in Proc. of Symposia in Applied Mathematics, vol. 47,
ed. I. Daubechies (AMS:  San Antonio)

Donoho, D.L., Johnstone, I.M. 1994, J. Amer. Stat. Asso., in press

Emslie, A.G. \& Horack, J.M.  1994, ApJ, in press.

Fenimore, E.E., et al.  1993, Nature, 366, 40

Fishman, G.J., et al.  1994, ApJ Suppl, in press

Horack, J.M., Emslie, A.G., \& Hartmann, D.  1994, ApJ, submitted

\hangindent=20pt
Hurley, K.  1991, in Gamma-Ray Bursts, AIP Conf. Proc. 265, eds. W.S.
Paciesas \& G.J.

Fishman (AIP: New York, 1991), p. 3

\hangindent=20pt
Klebesadel, R.W.  1990, in Gamma-Ray Bursts, eds. C. Ho, R.I. Epstein, \&
E.E. Fenimore (Cambridge Univ. Press: Cambridge), p. 161

\hangindent=20pt
Kouveliotou, C., Meegan, C.A., Fishman, G.J., Bhat, N.P., Briggs, M.S.,
Koshut, T.M., Paciesas, W.S., \& Pendleton, G.N.  1993, ApJ, 413, L101

\hangindent=20pt
Mao, S. \& Paczynski, B.  1992, ApJ, 388, L45

Mazets, E.P., et al.  1981, Ap. Space Sci., 80, 3

\hangindent=20pt
McCrea, W.H.  1972, in External Galaxies and Quasi Stellar Objects, ed.
D.S. Evans, IAU, p 283

\hangindent=20pt
Meegan, C.A., et al.  1994, The Second BATSE Burst Catalog, Compton
Observatory Science Support Center

\hangindent=20pt
Mitrofanov, I., et al.  1994, Gamma-Ray Bursts - Huntsville, AL 1993, AIP
307, in press

Nemiroff, R.J., et al.  1994, ApJ, in press

Noerdlinger, P.  1966, Ap.J., 143, L1004

Norris, J.P., Cline, T.L., Desai, U.D., \& Teegarden, B.J.  1984, Nature,
434, 308

Norris, J.P., et al.  1994a, Ap.J., 424, 540

\hangindent=20pt
Norris, J.P., Nemiroff, R.J., Davis, S.P., Kouveliotou, C., Fishman, G.J.,
Meegan, C.A., \& Paciesas, W.S.  1994b, in Gamma-Ray Bursts - Huntsville,
AL 1993, AIP 307, in press

Paczynski, B.  1992, Nature, 355, 521

Piran, T.  1992, ApJ, 389, L45

\hangindent=20pt
Press, W.H., Teukolsky, S.A., Vetterling, W.T., \& Flannery, B.P.  1992,
Numerical Recipes in Fortran (Cambridge:  Cambridge University Press), p.
614

\hangindent=20pt
Wickramasinghe, W.A.D.T., Nemiroff, R.J., Norris, J.P., Kouveliotou, C.,
Fishman, G.J., Meegan, C.A., Wilson, R.B., \& Paciesas, W.S.  1993, ApJ,
411, L55

Wijers, R.A.M.J. \& Paczynski, B.  1994, ApJ, in press

Yi, I., \& Mao, S.  1994, preprint

}
\vfill\eject

\cl{\bf FIGURE CAPTIONS }
\hangindent=0pt
\baselineskip=18pt
\parindent=0pt

\bigskip

{\bf Fig. 1.} - Duration distributions for bright BATSE bursts.  T90 (a)
and T50 (b) durations are measured from 5\% to 95\%, and 25\% to 75\%,
respectively, of integrated counts above fitted background in the energy
range $>$ 25 keV. Solid line:  with $S/N$ levels rendered equal (to that of
the dimmest burst analyzed) and with high and low pass filters applied.
Dashed line:  with original signal-to-noise levels and without application
of the filter procedure.

\bigskip

{\bf Fig. 2.} - T90 (a) and T50 (b) duration distributions for bright
(solid), dim (dashed), and dimmest (dotted) BATSE burst samples.  $S/N$
levels are rendered uniform (Method 1, see text).  Median duration from 11
noise realizations per burst is plotted.  Bursts with apparent total
durations less than 1.5 s are not included.

\bigskip

{\bf Fig. 3.} - Same as Figure 2, but duration analysis performed with
original $S/N$ levels preserved (Method 2).

\bigskip

{\bf Fig. 4.} - (a) Unbinned Kolmogorov-Smirnov test between bright and dim
duration distributions for Method 1.  Durations of bright bursts were
multiplied by candidate time-dilation factors and compared to the durations
of the two dim samples combined.  The solid and dot-dashed lines illustrate
probability as function of dilation factor for T90 and T50 durations,
respectively, that dim and time-dilated bright durations were drawn from
the same parent distribution.  Best time-dilation factors are 1.9 (T90) and
2.2 (T50).  (b) Same as (a) but for Method 2.  Because the distribution
widths measured with Method 2 are more nearly equal, the best time-dilation
factors (2.1 for T90 and 2.4 for T50) result in higher probabilities that
the dim and bright samples are drawn from the same population.

\vfill\eject
\end